\documentclass[aip,graphicx]{revtex4-1}

\usepackage{graphicx}

\draft 

\begin{document}





\title{Angle-dependent bandgap engineering in gated graphene superlattices}

\author{H. Garc\'ia-Cervantes}

\author{L. M. Gaggero-Sager}

\author{O. Sotolongo-Costa}

\affiliation{Facultad de Ciencias, Universidad Aut\'onoma del Estado de Morelos, Av. Universidad, Col. Chamilpa 62209 Cuernavaca, Morelos, M\'exico.}

\author{G. G. Naumis}

\affiliation{Instituto F\'isica, Depto. de F\'isica-Qu\'imica, Universidad Nacional Aut\'onoma de M\'exico (UNAM). Apdo. Postal 20-364, 01000, M\'exico D.F., M\'exico.}

\author{I. Rodr\'iguez-Vargas}

\affiliation{Unidad Acad\'emica de F\'isica, Universidad Aut\'onoma de Zacatecas, Calzada Solidaridad Esquina Con Paseo La Bufa S/N, 98060 Zacatecas, Zac., M\'exico.}

\date{\today}

\begin{abstract}
Graphene Superlattices (GSs) have attracted a lot of attention due to its peculiar properties as well as its possible technological implications. Among these characteristics we can mention: the extra Dirac points in the dispersion relation and the highly anisotropic propagation of the charge carriers. However, despite the intense research that is carried out in GSs, so far there is no report about the angular dependence of the Transmission Gap (TG) in GSs. Here, we report the dependence of TG as a function of the angle of the incident Dirac electrons in a rather simple Electrostatic GS (EGS). Our results show that the angular dependence of the TG is intricate, since for moderated angles the dependence is parabolic, while for large angles an exponential dependence is registered.  We also find that the TG can be modulated from meV to eV, by changing the structural parameters of the GS. These characteristics open the possibility for an angle-dependent bandgap engineering in graphene.  
\end{abstract}


\maketitle 

\begin{figure}[h]
\includegraphics[scale=0.3]{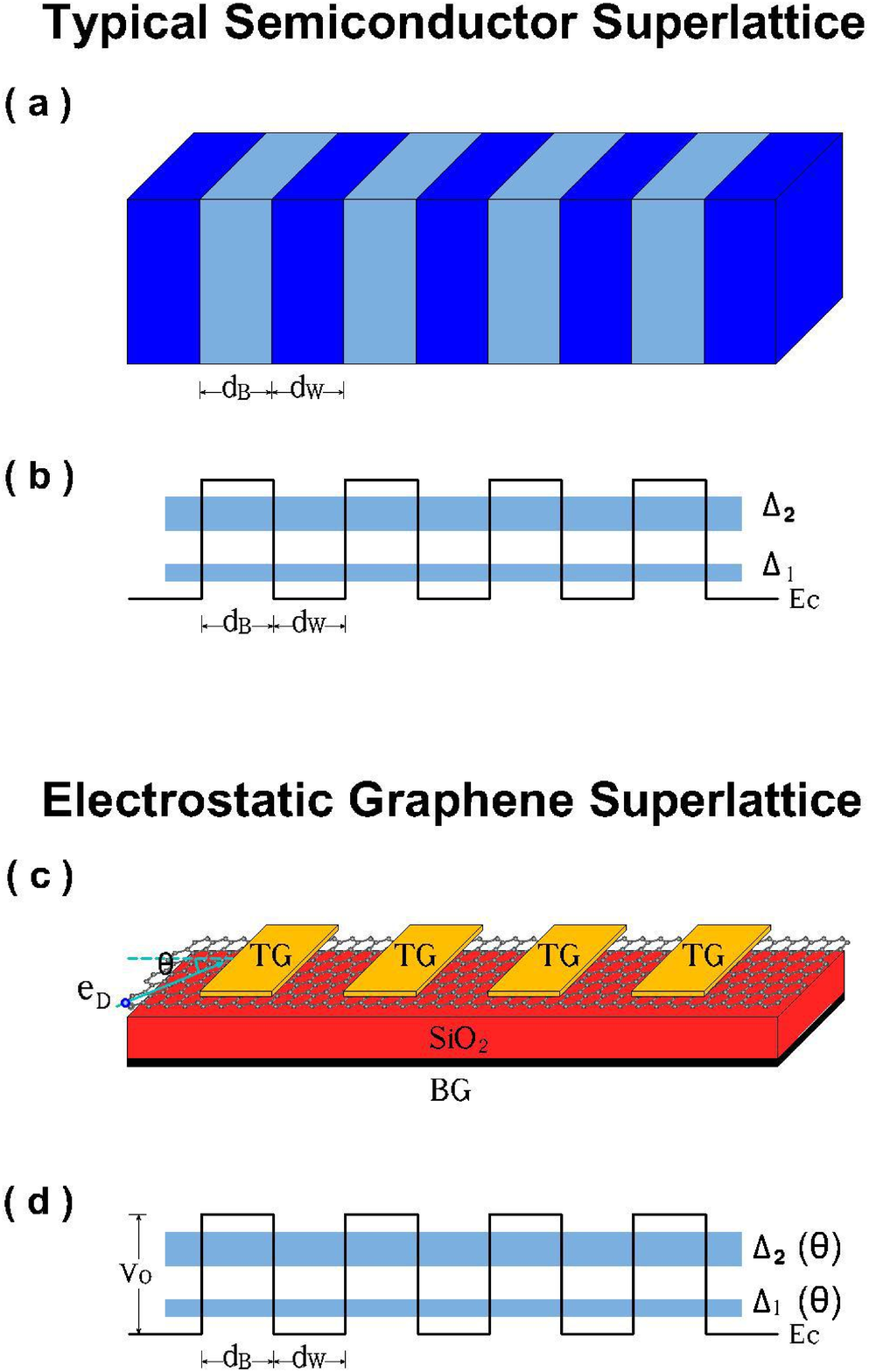}
\caption{\label{Fig1} Schematic representation of Typical Semiconductor and Electrostatic Graphene Superlattices. The cross-sections of these superlattices are shown in (a) and (c), while in (b) and (d) the conduction band-edge profiles and the energy minibands ($\Delta_1$ and $\Delta_2$) of them are depicted. $d_B$, $d_W$ and $V_0$ represent the widths of barriers and wells, and the height of the barriers, respectively. Aside to the structural differences, the main physical difference is that energy minibands and gaps in the case of Electrostatic Graphene Superlattices depend on the angle of incidence. } 
\end{figure}

Bandgap engineering or band structure engineering is a term coined in the late eighties to refer to a powerful technique for the design of new semiconductor materials and devices.\cite{FCapasso1987Science} This technique is based on the ability to modify the energy bands arbitrarily and to tailor them for a specific application. Among the typical tools to achieve bandgap engineering we can mention: doping, materials with variable gap, band discontinuities and superlattices. These tools can be used alone or in combination to obtain a particular band structure. For instance, superlattices are artificial periodic structures that can be created by alternating semiconductors with disimilar bandgaps, see Fig. \ref{Fig1}a.  The difference of bandgaps of the constituent materials as well as the super-periodicity of the structure turn out in a periodic band-edge profile of barriers and wells, giving rise to allowed and forbidden energy bands, commonly known as energy minibands and gaps respectively, Fig. \ref{Fig1}b.  Semiconductor superlattices present a plenty of physical effects that can be exploited technologically, among them we can find excitonic effects, miniband transport, Wannier–Stark localization, Bloch oscillations, electric field domains, resonant and sequential tunnelling.\cite{HGrahn1995Book} Specifically, superlattices can be used as injection medium or as active region in the well-known Quantum Cascade Laser.\cite{JFaist1994Science}

Within this context, bandgap engineering in graphene is not the exception, and from the very beginning of the discovery of graphene\cite{KNovoselov2004Science,KNovoselov2005Nature,YZhang2005Nature} the scientific community has tried to figure out how to create and modify a bandgap in graphene.\cite{YSon2006PRL,MHan2007PRL,XWang2008PRL,SZhou2007NatMater,GGiovannetti2007PRB,XPeng2008NL,
ECastro2007PRL,JOostinga2008NatMater,RBalog2010NatMater,DHaberer2010NL,JBai2010NatNanotech,XLiang2010NL,
GGui2010PRB,FGuinea2010NatPhys} This in part due to the gapless dispersion relation of graphene\cite{PWallace1947PR,AHCNeto2009RMP} as well as a natural technique to tailor the band structure of graphene-based structures for a specific application. Among the different approaches to create a bandgap we can mention graphene nanoribbons,\cite{YSon2006PRL,MHan2007PRL,XWang2008PRL} epitaxial substrates,\cite{SZhou2007NatMater,GGiovannetti2007PRB,XPeng2008NL} biased bilayer graphene,\cite{ECastro2007PRL,JOostinga2008NatMater} hydrogen absorption,\cite{RBalog2010NatMater,DHaberer2010NL} graphene nanomesh,\cite{JBai2010NatNanotech,XLiang2010NL} and strain engineering.\cite{GGui2010PRB,FGuinea2010NatPhys} All these approaches are based on the aperture and modulation of a bandgap. However, as we above mention there are another possibilities, for example superlattices. Indeed, superlattices in graphene have been studied intensely in the past few years.\cite{CBai2007PRB,MBarbier2008PRB,CPark2008NatPhys,CPark2008PRL,CPark2008NL,MBarbier2009PRB,MBarbier2010PRB,
SRusponi2010PRL,PBurset2011PRB,MYankowitz2012NatPhys,LPonomarenko2013Nature,
BHunt2013Science,LDellAnna2009PRB,RBiswas2010JAP,MRMasir2010JPCM,LDellAnna2011PRB,XGuo2011APL,
GMaksimova2012PRB,MTitov2010PRB,FPellegrino2012PRB,HYan2013PRB,LChernozatonskii2007APL,MYang2010APL} Among the most important characteristic of graphene superlattices we can found: additional Dirac cones in the energy-dispersion relation and highly anisotropic propagation of charge carriers. The latter is quite interesting, since depending on the transversal wave vector or the angle of the electrons that impinging on the superlattice structure the propagation properties can be tuned readily. This opens the way for possible electron wave filters and laser devices. However, as a first step to propose this kind of devices, it is important to know in detail the angular dependence of the energy minibands and gaps, aspect that as far as we know is not reported up to date.

Here, we propose a bandgap engineering based on the angular dependence of the propagation of Dirac electrons in graphene superlattices. Our approach relies on the formation of transmission minibands and gaps at diferente energy scales depending on the angle of incidence of Dirac electrons. In particular, we show that the angular dependence of the TG is intricate, being parabolic for moderate angles and exponential for large ones. To this respect, TGs from meV to various eV can be obtained by angular selection as well as by tailoring the structural parameters of GSs (number of periods, height of barriers, and widths of barriers and wells). So, angle-dependent bandgap engineering gives the possibility to modify the energy minibands and gaps in GSs almost arbitrarily and to tailor them for a specific application.  

The system we are interested in is a rather simple electrostatic graphene superlattice. In Fig. \ref{Fig1}c we show a schematic representation of this structure, it consists of a graphene sheet sitting on a non-interacting substrate like SiO$_2$, a back gate, and top gates arranged periodically along the superlattice axis ($x$). In the left part of the structure, it is also shown an electron impinging at a certain angle with respect to the superlattices axis. We have chosen this structure because from the experimental standpoint is more reliable.\cite{NSander2009PRL,AYoung2009NatPhys} For instance, the energy of the incident electrons and the width and height of the potential barriers can be controlled by the doping of the substrate and the strength of the voltages applied to the back and top gates. The band-edge profile of this structure can be modelled as a periodic arrangement of potential barriers and wells, that in turn results in energy minibands and gaps, which strongly depend on the angle of the impinging electrons, see Fig. \ref{Fig1}d. The transmission properties of this system can be computed  straightforwardly using the transfer matrix approach. \cite{PYeh2005,CSoukoulis2008} The basic information needed to apply this methodology is the dispersion relation, wave vectors and wave functions in the barrier and well regions as well as in the semi-infinite left and right regions.\cite{IRVargas2012JAP,JABTorres2014SM} In the well and semi-infinite regions the dispersion relation and wave functions comes as:

\[
E=\pm \hbar v_F k,
\]

\noindent and 

\begin{equation}
\psi_{\pm}^k (x,y) = \frac{1}{\sqrt{2}} 
\left( 
\begin{array}{c}
1 \\
u_{\pm}
\end{array}
\right) e^{\pm ik_x x + ik_y y},
\end{equation}

\noindent where $v_F$ is the Fermi velocity, $k$ is the magnitud of the wave vector in these regions, $k_x$ and $k_y$ are the longitudinal and transversal components of $k$, and $u_{\pm}= \mbox{sign(E)} e^{\pm i\theta}$ the coefficients of the wave functions that depend on the angle of the impinging electrons, $\theta=\arctan (k_y/k_x)$. In the barrier regions these quantities come as:

\[
E= V_0 \pm \hbar v_F q,
\]

\noindent and 

\begin{equation}
\psi_{\pm}^q (x,y) = \frac{1}{\sqrt{2}} 
\left( 
\begin{array}{c}
1 \\
v_{\pm}
\end{array}
\right) e^{\pm iq_x x + iq_y y},
\end{equation}
 
\noindent where $V_0$ is the strength of the electrostatic potential, $q$ is the magnitud of the wave vector in the barrier regions, $q_x$ and $q_y$ are the components of $q$, and $v_{\pm}$ the coefficients of the wave functions.\cite{IRVargas2012JAP,JABTorres2014SM} Once these quantities are known, we can apply the continuity conditions of the wave function along the superlattice axis as well as the consevation of the transversal momentum($k_y=q_y$), and define the transmission probability in terms of the so-called transfer matrix, 

\begin{equation}
T(E,\theta)= \frac{1}{| M_{11} |^2},
\end{equation}

\noindent which depends on the transfer matrices of barriers and wells, and the number of periods as well, for more details see Refs. 50 and 51. 

\begin{figure}[h]
\includegraphics[scale=0.7]{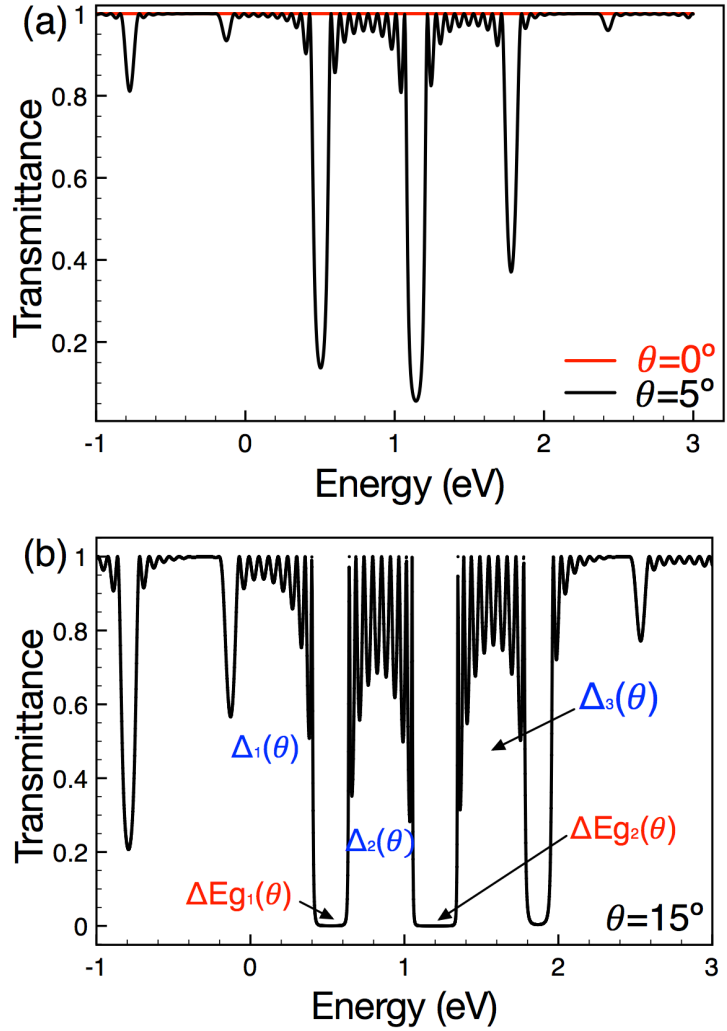}
\caption{\label{Fig2}Formation of energy minibands and gaps in EGSs as a function of the angle ($\theta$) of the impinging electrons. In (a) the transmittance for $\theta=0^{\circ}$ and $\theta=5^{\circ}$ is presented, while in (b) the transmission probability for $\theta=15^{\circ}$ is shown. We have denoted the energy minibands as $\Delta_1$, $\Delta_2$ and $\Delta_3$, and the energy gaps as $\Delta$Eg$_1$ and $\Delta$Eg$_2$.}
\end{figure}

\begin{figure}[h]
\includegraphics[scale=0.7]{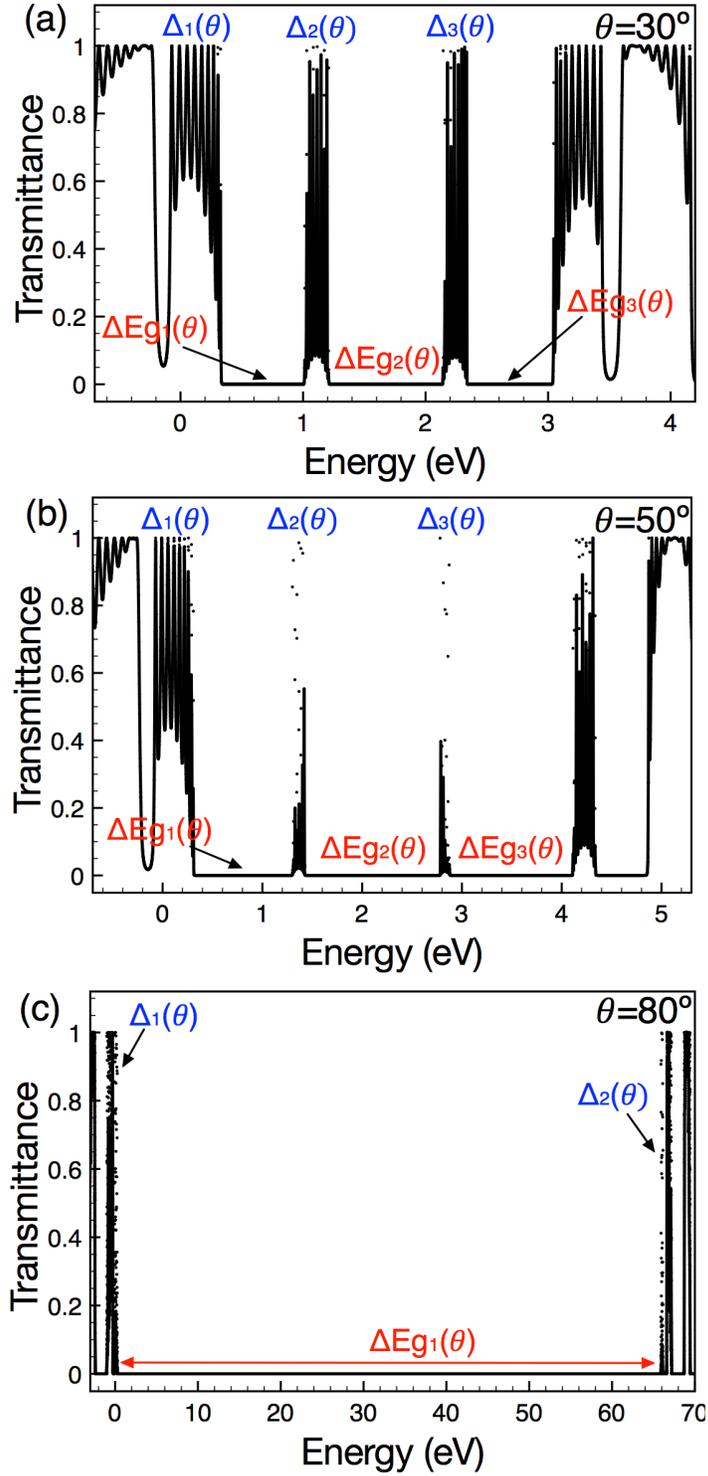}
\caption{\label{Fig3} Evolution of the energy minibands and gaps in EGSs for different angles of incidence: (a) $30^{\circ}$, (b) $50^{\circ}$ and (c) $80^{\circ}$, respectively. In general, the energy minibands become narrower, while the energy gaps get larger as the angle of incidence increases. Even more, a dramatic change is presented in (c), where the first transmission gap changes hugely.}
\end{figure}

\begin{figure}[h]
\includegraphics[scale=0.45]{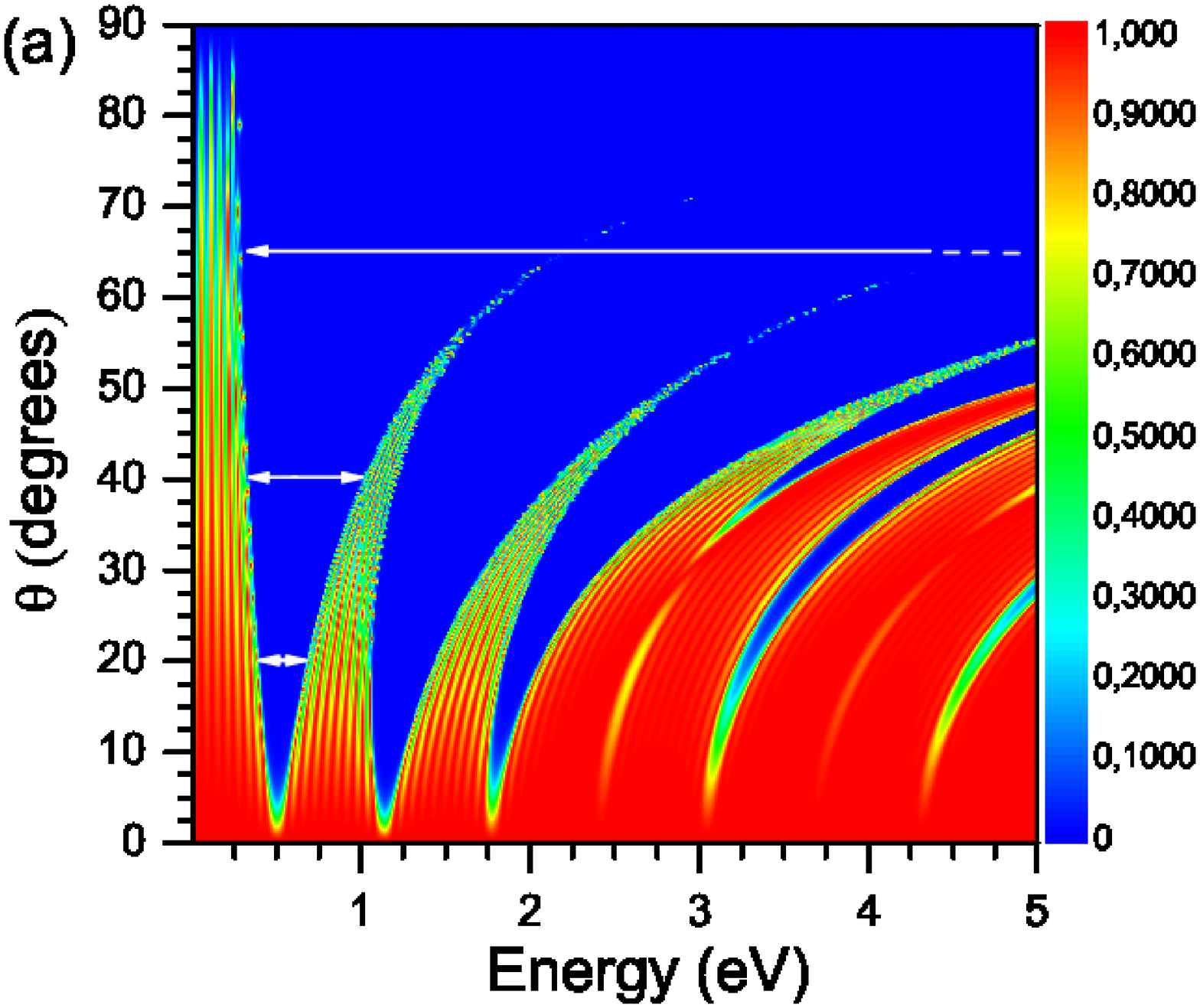}
\includegraphics[scale=0.45]{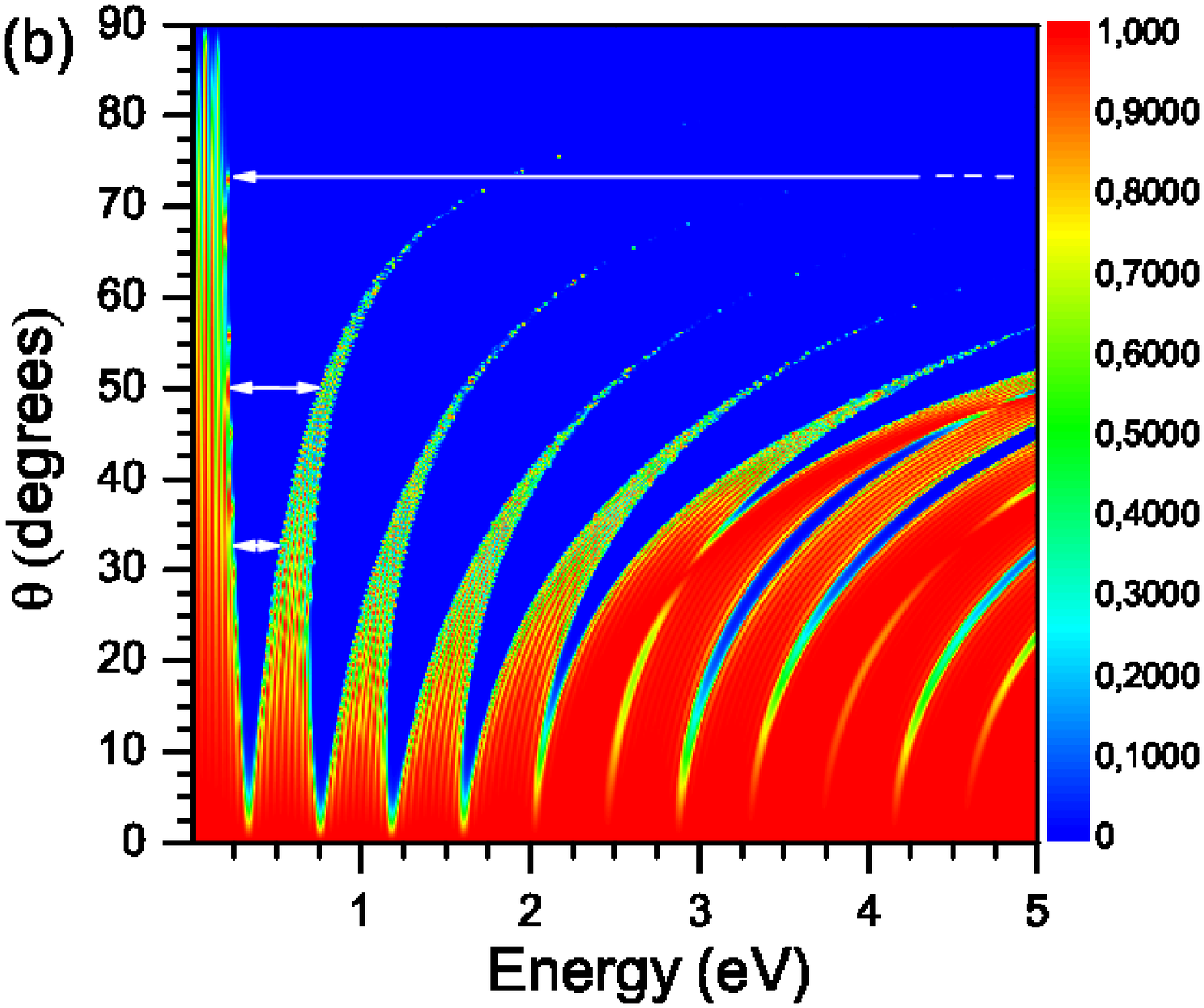}
\caption{\label{Fig4} Contour plots of the transmittance as a function of the energy and angle of the incident electrons: (a) Transmission contour of EGS for superlattice parameters: $N=10$, $d_B=d_W=10a$ and $V_0=1$ eV, (b) The same as in (a), but with double of the size of the width of the quantum wells, $d_W=20a$. White arrows help us to guide the evolution of the first transmission gap, and specifically to indicate that at certain angle there is a dramatic change due to the occlusion of various transmission minibands.}
\end{figure}

At first, we want to discuss the formation of energy minibands and gaps as a function of the angle of incidence. In Fig. \ref{Fig2} we show the transmission probability as a function of the energy for three different angles. Specifically, Fig. \ref{Fig2}a depicts the transmittance for normal incidence and $\theta=5^{\circ}$, solid-red and solid-black lines respectively, while Fig. \ref{Fig2}b shows the case of $\theta=15^{\circ}$. The number of periods, the widths of barriers and wells, and the height of the barriers are $N=10$, $d_B=10a$, $d_W=10a$ and $V_0=1.0$ eV, respectively. These parameters will remain fixed, virtually, throughout the study. As we can see Klein tunneling (perfect transmission) prevents the formation of energy minibands and gaps for normal incidence, $\theta=0^{\circ}$. Once the angle of incidence is different from zero, transmission minibands and gaps start to develop. For small angles, as in the case of $5^{\circ}$, we have pseudo minibands and gaps,\cite{YXu2015PhysicaB} since them are not well defined yet. By increasing systematically $\theta$ we will find that the mentioned pseudo minibands and gaps become well defined ones. This process takes place in staircase fashion, due to not all transmission minibands and gaps become well defined at the same angle. For instance, in the case of $\theta=15^{\circ}$ we can see that the first and second gaps are almost well defined as well as the second and third minibands, however what we labeled as first miniband in reality is a pseudo miniband, since in the low energy side of it there is not a well defined gap. Even more, this miniband is shared among electrons and holes, positive and negative energies, respectively. By increasing the angle of incidence to $30^{\circ}$ and $50^{\circ}$ we can see that the widths of the energy minibands diminish, while the corresponding ones to the transmission gaps increase, see Fig. \ref{Fig3}a and b. In the specific case of the minibands, we can also notice that the number of resonances within them diminish as well. This reduction is quite important, since for example a particular miniband will occlude in a specific angle, and consequently the corresponding transmission gap will increase substantially. For instance, in Fig. \ref{Fig3}c we can see that the first transmission gap is huge as a consequence of the collapsing of various minibands, or in other words the second miniband is at quite different energy scale, due to the occlusion of various transmission minibands. We can have a better perspective if we taking into account the contour plot of the transmittance as a function of both the energy and angle of the incident electrons, see Fig. \ref{Fig4}a. In fact, the contour tells us that the energy minibands and gaps are formed in a staircase fashion, the energy minibands have a semi-circular form, akin to a whisker, and are wider for small angles and virtually disappear for large angles. In addition, as the energy increases these whiskers bend, and the bending is steeper for higher minibands, as a consequence higher minibands occlude at lesser angles than lower minibands. Other important characteristics that we can find in EGSs are: 1) the number of resonances within a miniband is proportional to the number of periods in the superlattice, 2) by increasing the number of periods it is also possible to obtain well defined minibands and gaps irrespective of the angle of incidence,\cite{YXu2015PhysicaB} except normal incidence, 3) by changing the widths of the barriers and wells as well as the height of the barriers it is possible to tune the number of minibands and gaps, and their energy location, see Fig. \ref{Fig4}b.  

\begin{figure}[h]
\includegraphics[scale=0.7]{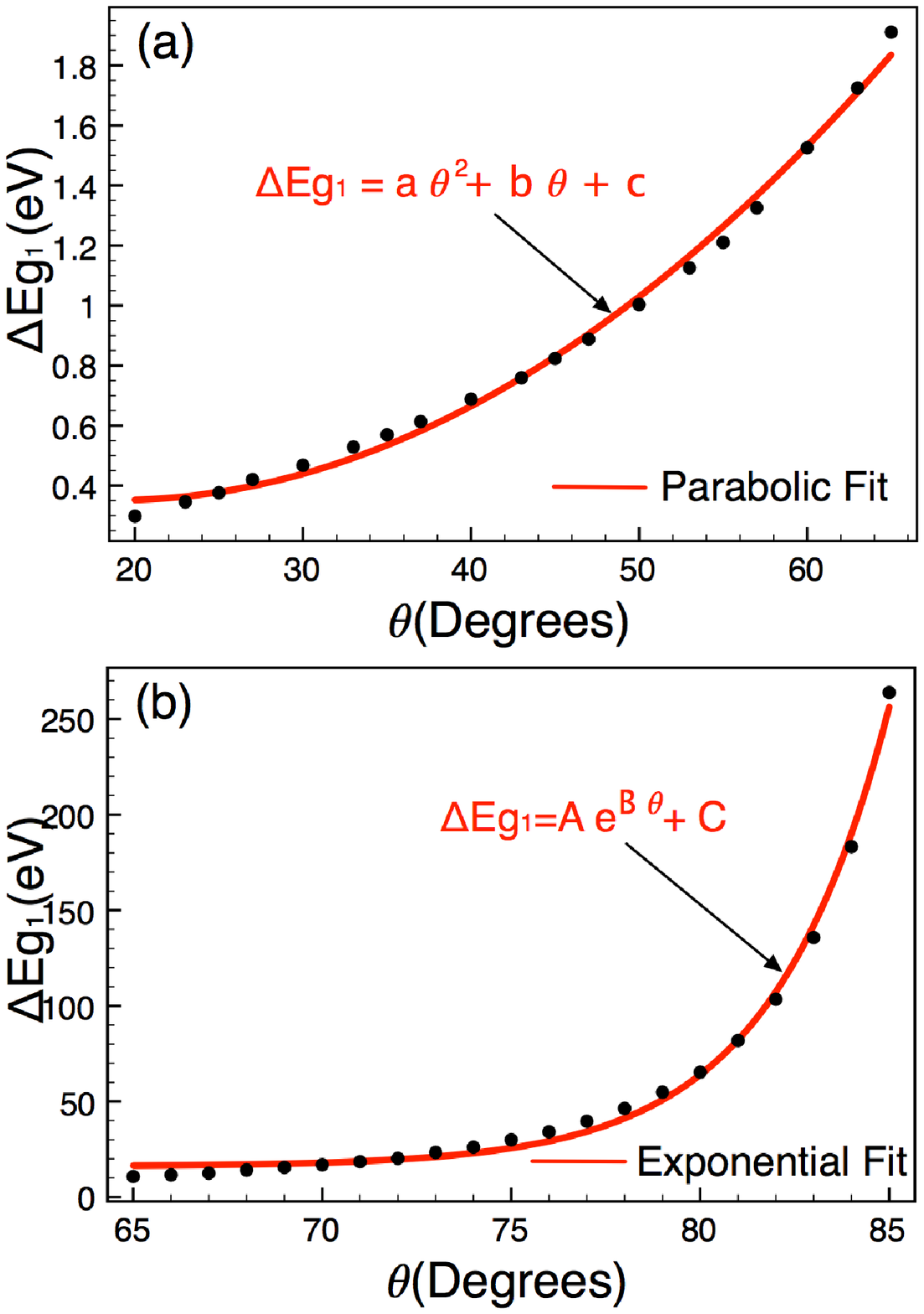}
\caption{\label{Fig5} Angular dependence of the first transmission gap $\Delta$Eg$_1$ in EGSs. (a) In the angular range of $20^{\circ}$ to $65^{\circ}$ a parabolic dependence is presented, while (b) an exponential dependence arises from $65^{\circ}$ to $90^{\circ}$	. In particular, the parabolic and exponential dependence give the possibility to modulate $\Delta$Eg$_1$ from meV to eV. The solid-red lines correspond to the parabolic and exponential fit, respectively. The Adj. R-Square is 0.99418 and 0.99507 for the parabolic and exponential fit, indicating that the fitting is quite good in both cases.}
\end{figure}

Within this context, a natural question arises, what is the angular dependence of the transmission minibands and gaps?. In order to answer this question, and without lose of generality, we will focus on what we labeled as first transmission gap. As we can see from the transmission contours, it is possible to infer that there are two angular regions for the first transmission gap due to the dramatic change that is presented about $60^{\circ}$-$70^{\circ}$, see Fig. \ref{Fig4}. Indeed, we find a parabolic dependence in the range of $20^{\circ}$ to $65^{\circ}$, while in the interval of $65^{\circ}$ to $90^{\circ}$ an exponential dependence is presented, see Fig. \ref{Fig5}. Excellent parabolic and exponential fits were found, see solid-red lines in Fig. \ref{Fig5}. For instance, the Adj. R-Square is 0.99418 and 0.99507 for the parabolic and exponential fit, which means that the fitting is quite good. The fitting parameters in the parabolic region are $a=6.93 \times 10^{-4}$ eV, $b=-0.026$ eV and $c=0.594$ eV, while in the exponential region we have $A=2.99 \times 10^{-10}$ eV, $B=0.322^{\circ{-1}}$ and $C=16.09$ eV. It is also important to highlight that choosing appropriately the angle of incidence, the first transmission gap can be tune from meV (small angles) to various eV (large angles). This opens the possibility of a bandgap engineering with a tremendous energy range of modulation, notice the vertical scale in Fig. \ref{Fig5}. This is a unique property presented by EGSs, since to modulate the bandgap in traditional semiconductor superlattices, it is mandatory to change mainly the constituent materials. Here, on the contrary, it is required to have full control of the angle of incidence, or more realistically speaking, total control over a narrow angular interval. To this respect, we can expect that with the current experimental technologies this requirement be totally affordable. In fact, recently, important steps have been made in order to find out the angular contribution of Dirac electrons to the transport properties in graphene-based devices.\cite{ARahman2015APL,RSajjad2013ACSNano,RSajjad2012PRB,SSutar2012NanoLett} Specifically, some groups have been able to discriminate the angular contribution by using tilting metallic electrodes (top gates) in single-barrier graphene structures. Finally, it is important to remark that in order to angle-dependent bandgap engineering be realistic and reliable other relevant issues have to be addressed. For instance, in the case of laser devices, studies about tunneling rates, relaxation times, tunneling escape probabilities, emission lifetimes and radiative efficiency are needed in order to elucidate the conditions for population inversion. 

In summary, we have proposed an angle-dependent bandgap engineering in graphene. This approach is based on the angular dependence of the propagation of Dirac electrons in EGSs. Specifically, transmission minibands and gaps can be formed at different energy ranges depending on the angle of the incident electrons. In the case of the transmission gaps, the main gap shows a parabolic and exponential angular dependence for moderate and large incident angles, respectively. The transmission gap can be tuned from meV to eV by angular selection as well as by changing the structural parameters of GSs. So, in principle GSs offer a tremendous possibility to design and develop devices based on angle-dependent bandgap engineering.

\pagebreak


\begin{thebibliography}{99}

\bibitem{FCapasso1987Science} F. Capasso, Science \textbf{235}, 172 (1987). 

\bibitem{HGrahn1995Book} H.T. Grahn, Semiconductor Superlattices: Growth and Electronic Properties, World Scientific, 1995.

\bibitem{JFaist1994Science} J. Faist, F. Capasso, D. L. Sivco, C. Sirtori, A. L. Hutchinson, A. Y. Cho, Science \textbf{264}, 553 (1994). 

\bibitem{KNovoselov2004Science} K. S. Novoselov, A. K. Geim, S. V. Mrozov, D. Jiang, Y. Zhang, S. V. 
Dubonos, I. V. Grigorieva, and A. A. Firsov, Science \textbf{306}, 666 (2004).

\bibitem{KNovoselov2005Nature} K. S. Novoselov, A. K. Geim, S. V. Morozov, D. Jiang, M. I. Katsnelson, I. V. Grigorieva, S. V. Dubonos, and A. A. Firsov, Nature \textbf{438}, 197 (2005).

\bibitem{YZhang2005Nature} Y. Zhang, Y.-W. Tan, H. L. Stormer, and P. Kim, Nature \textbf{438}, 201 (2005).


\bibitem{YSon2006PRL} Y.-W. Son, M. L. Cohen, and S. G. Louie, Phys. Rev. Lett. \textbf{97}, 210803 (2006). 

\bibitem{MHan2007PRL} M. Y. Han, B. \"Ozyilmaz, Y. Zhang, and P. Kim, Phys. Rev. Lett. \textbf{98}, 206805 (2007). 

\bibitem{XWang2008PRL} X. Wang, Y. Ouyang, X. Li, H. Wang, J. Guo, and H. Dai, Phys. Rev. Lett. \textbf{100}, 206803 (2008). 

\bibitem{SZhou2007NatMater} S.Y. Zhou, G.-H. Gweon, A.V. Fedorov, P.N. First, W.A. de Heer, D.-H. Lee, F. Guinea, A.H. Castro-Neto, A. Lanzara, Nat. Mater. \textbf{6}, 770 (2007).

\bibitem{GGiovannetti2007PRB} G. Giovannetti, P. A. Khomyakov, G. Brocks, P. J. Kelly, and J. van den Brink, Phys. Rev. B \textbf{76}, 073103 (2007).

\bibitem{XPeng2008NL} X. Peng and R. Ahuja, Nano Lett. \textbf{8}, 4464 (2008). 

\bibitem{ECastro2007PRL} E. V. Castro, K. S. Novoselov, S. V. Morozov, N. M. R. Peres, J. M. B. Lopes dos Santos, Johan Nilsson, F. Guinea, A. K. Geim, and A. H. Castro Neto, Phys. Rev. Lett. \textbf{99}, 216802 (2007). 

\bibitem{JOostinga2008NatMater} J. B. Oostinga, H. B. Heersche, X. Liu, A. F. Morpurgo, and L. M. K. Vandersypen, Nat. Mater. \textbf{7}, 151 (2008). 

\bibitem{RBalog2010NatMater} R. Balog et al., Nat. Mater. \textbf{9}, 315 (2010). 

\bibitem{DHaberer2010NL} D. Haberer, D. V. Vyalikh, S. Taioli, B. Dora, M. Farjam, J. Fink, D. Marchenko, T. Pichler,O K. Ziegler, S. Simonucci, M. S. Dresselhaus, M. Knupfer, B. B\"chner, and A. Gr\"uneis, Nano Lett. \textbf{10}, 3360 (2010). 

\bibitem{JBai2010NatNanotech} J. Bai, X. Zhong, S. Jiang, Y. Huang, and X. Duan, Nat. Nanotechnol.  \textbf{5}, 190 (2010). 

\bibitem{XLiang2010NL} X. Liang, Y.-S. Jung, S. Wu, A. Ismach, D. L. Olynick, S. Cabrini, and J. Bokor, Nano Lett. \textbf{10}, 2454 (2010). 

\bibitem{GGui2010PRB} G. Gui, J. Li, and J. Zhong, Phys. Rev. B \textbf{78}, 075435 (2008). 

\bibitem{FGuinea2010NatPhys} F. Guinea, M. I. Katsnelson and A. K. Geim, Nat. Phys. \textbf{6}, 30 (2010). 

\bibitem{PWallace1947PR} P. R. Wallace, Phys. Rev. \textbf{71}, 622 (1947).

\bibitem{AHCNeto2009RMP} A. H. Castro Neto, F. Guinea, N. M. R. Peres, K. S. Novoselov and A. K. Geim, Rev. Mod. Phys. \textbf{81}, 109 (2009). 

\bibitem{CBai2007PRB} C. Bai and X. Zhang, Phys. Rev. B \textbf{76}, 75430 (2007).

\bibitem{MBarbier2008PRB} M. Barbier, F.M. Peeters, P. Vasilopoulos, J.M. Pereira Jr., Phys. Rev. B \textbf{77}, 115446 (2008).

\bibitem{CPark2008NatPhys} C.-H. Park, L. Yang, Y.-W. Son, M. L. Cohen, and S. G. Louie, Nat. Phys. \textbf{4}, 213 (2008). 

\bibitem{CPark2008PRL} C.-H. Park, Li Yang, Y.-W. Son, M. L. Cohen, and S. G. Louie, Phys. Rev. Lett. \textbf{101}, 126804 (2008).

\bibitem{CPark2008NL} C.-H. Park, Y.-W. Son, L. Yang, M.L. Cohen, S.G. Louie, Nano Lett. \textbf{8}, 2920 (2008).

\bibitem{MBarbier2009PRB} M. Barbier, P. Vasilopoulos, F.M. Peeters, Phys. Rev. B \textbf{80}, 205415 (2009).

\bibitem{MBarbier2010PRB} M. Barbier, P. Vasilopoulos, F.M. Peeters, Phys. Rev. B 81 (2010) 075438.

\bibitem{SRusponi2010PRL} S. Rusponi, M. Papagno, P. Moras, S. Vlaic, M. Etzkorn, P. M. Sheverdyaeva, D. Pacil\'e, H. Brune, and C. Carbone, Phys. Rev. Lett. \textbf{105}, 246803 (2010). 

\bibitem{PBurset2011PRB} P. Burset, A. Levy Yeyati, L. Brey, H.A. Fertig, Phys. Rev. B 83, 195434 (2011).

\bibitem{MYankowitz2012NatPhys} M. Yankowitz,	J. Xue,	D. Cormode, J. D. Sanchez-Yamagishi, K. Watanabe,	T. Taniguchi,	P. Jarillo-Herrero, P. Jacquod and B. J. LeRoy, Nat. Phys. \textbf{8}, 382 (2012). 

\bibitem{LPonomarenko2013Nature} L. A. Ponomarenko,	R. V. Gorbachev,	G. L. Yu,	D. C. Elias, R. Jalil, A. A. Patel,	A. Mishchenko,	 A. S. Mayorov,	C. R. Woods, J. R. Wallbank,	M. Mucha-Kruczynski,	B. A. Piot, M. Potemski, I. V. Grigorieva,	K. S. Novoselov, F. Guinea, V. I. Fal’ko and A. K. Geim, Nature \textbf{497}, 594 (2013). 

\bibitem{BHunt2013Science} B. Hunt, J. D. Sanchez-Yamagishi, A. F. Young1, M. Yankowitz, B. J. LeRoy, K. Watanabe, T. Taniguchi, P. Moon4, M. Koshino, P. Jarillo-Herrero, R. C. Ashoori, Science \textbf{340}, 1427 (2013). 

\bibitem{LDellAnna2009PRB} L. Dell'Anna, A. De Martino, Phys. Rev. B \textbf{79}, 045420 (2009).

\bibitem{RBiswas2010JAP} R. Biswas, A. Biswas, N. Hui, C. Sinha, J. Appl. Phys. 108, 043708 (2010).

\bibitem{MRMasir2010JPCM} M. Ramezani Masir, P. Vasilopoulos, F.M. Peeters, J. Phys.: Condens. Matter \textbf{22}, 465302 (2010).

\bibitem{LDellAnna2011PRB} L. Dell'Anna, A. De Martino, Phys. Rev. B \textbf{83}, 155449 (2011).

\bibitem{XGuo2011APL} X.-X. Guo, D. Liu, Y.-X. Li, Appl. Phys. Lett. \textbf{98}, 242101 (2011).

\bibitem{GMaksimova2012PRB} G.M. Maksimova, E.S. Azarova, A.V. Telezhnikov, V.A. Burdov, Phys. Rev. B \textbf{86},  205422 (2012).

\bibitem{MTitov2010PRB} S. Gattenl\"ohner, W. Beizig, M. Titov, Phys. Rev. B \textbf{82}, 155417 (2010).

\bibitem{FPellegrino2012PRB} F. M. D. Pellegrino, G. G. N. Angilella, R. Pucci, Phys. Rev. B \textbf{85}, 195409 (2012).

\bibitem{HYan2013PRB} H. Yan, Z.-D. Chu, W. Yan, M. Liu, L. Meng, M. Yang, Y. Fan, J. Wang, R.-F. Dou, Y. Zhang, Z. Liu, J.-C. Nie, L. He, Phys. Rev. B \textbf{87}, 075405 (2013).

\bibitem{LChernozatonskii2007APL} L.A. Chernozatonskii, P.B. Sorokin, J.W. Brüning, Appl. Phys. Lett. \textbf{91}, 183103 (2007).

\bibitem{MYang2010APL} M. Yang, A. Nurbawono, C. Zhang, Y.P. Feng, Ariando, Appl. Phys. Lett. \textbf{96}, 193115 (2010).

\bibitem{NSander2009PRL} N. Stander, B. Huard, D. Goldhaber-Gordon, Phys. Rev. Lett. \textbf{102}, 026807 (2009).

\bibitem{AYoung2009NatPhys} A. F. Young, P. Kim, Nat. Phys. \textbf{5}, 222 (2009).

\bibitem{PYeh2005} P. Yeh, \textit{Optical Waves in Layered Media} (Wiley-Interscience, 2005).

\bibitem{CSoukoulis2008} P. Markos and C. M. Soukoulis, \textit{Wave Propagation: From Electrons to Photonic Crystals and Left-Handed Materials} (Princeton University Press, 2008). 

\bibitem{IRVargas2012JAP} I. Rodr\'iguez-Vargas, J. Madrigal-Melchor and O. Oubram, J. Appl. Phys. \textbf{112}, 073711 (2012). 

\bibitem{JABTorres2014SM} J. A. Briones-Torres, J. Madrigal-Melchor, J. C. mart\'inez-Orozco and I. Rodr\'iguez-Vargas, Superlattices and Microstructures \textbf{73}, 98 (2014). 

\bibitem{YXu2015PhysicaB} Y. Xu, Y. He and Y. Yang, Physica B \textbf{457}, 188 (2015). 

\bibitem{ARahman2015APL} A. Rahman, J. W. Guikema, M. Hassan, and N. Markovic, Appl. Phys. Lett. \textbf{103}, 013112 (2015). 

\bibitem{RSajjad2013ACSNano}  R. N. Sajjad and A. W. Ghosh, ACS nano \textbf{7}, 9808 (2013).

\bibitem{RSajjad2012PRB} R. N. Sajjad, S. Sutar, J. U. Lee, and A. W. Ghosh, Phys. Rev. Lett. \textbf{86}, 155412 (2012). 

\bibitem{SSutar2012NanoLett} S. Sutar, E. S. Comfort, J. Liu, T. Taniguchi, K. Watanabe, and J. U. Lee, Nano Lett. \textbf{12}, 4460 (2012). 

\end{thebibliography}
\end{document}